\documentclass[english,10pt,final,twocolumn]{elsarticle}
\usepackage[T1]{fontenc}
\usepackage[latin9]{inputenc}
\usepackage{booktabs}
\usepackage{pdfpages}
\usepackage{amsmath}
\usepackage{graphicx}

\makeatletter

\usepackage{braket}
\usepackage{mathrsfs}
\usepackage{slashed}
\usepackage{bm}
\usepackage{datetime}
\usepackage{siunitx}

\DeclareSIUnit[number-unit-product = {}]\clight{c}
\DeclareSIUnit\eVperc{\eV\per\clight}
\DeclareSIUnit\GeVpercs{\giga\eV\squared\per\clight\squared}
\DeclareSIUnit\MeVpercs{\mega\eV\per\clight\squared}
\journal{Physics Letters B}
\biboptions{numbers,sort&compress}
\usepackage[left=1.6cm,right=1.04cm,top=1.65cm,bottom=1.65cm,columnsep=25pt]{geometry}
\usepackage{lineno}

\usepackage{hyperref}  
\hypersetup{breaklinks = true, colorlinks = true, allcolors = magenta}
\usepackage{setspace}
\usepackage{graphicx}
\usepackage{wrapfig}

\makeatother

\usepackage{babel}
\begin{document}

\begin{frontmatter}{}

\title{Measurement of polarization-transfer to bound protons in carbon \\
and its virtuality dependence}

\author[TAU]{D.~Izraeli\fnref{eq}\corref{cor2}}

\ead{davidizraeli@post.tau.ac.il}

\author[JSI]{T.~Brecelj\fnref{eq}}

\author[Mainz]{P.~Achenbach}

\author[TAU]{A.~Ashkenazi}

\author[Mainz]{R.~B\"ohm}

\author[TAU]{E.O.~Cohen}

\author[Mainz]{M.O.~Distler}

\author[Mainz]{A.~Esser}

\author[Rutgers]{R.~Gilman}

\author[JSI]{T.~Kolar}

\author[TAU,nrc]{I.~Korover}

\author[TAU]{J.~Lichtenstadt}

\author[TAU,soreq]{I.~Mardor}

\author[Mainz]{H.~Merkel}

\author[JSI,Mainz]{M.~Mihovilovi\v{c} }

\author[Mainz]{U.~M\"uller}

\author[TAU]{M.~Olivenboim}

\author[TAU]{E.~Piasetzky}

\author[huji]{G.~Ron}

\author[Mainz]{B.S.~Schlimme}

\author[Mainz]{M.~Schoth}

\author[Mainz]{C.~Sfienti}

\author[UL,JSI]{S.~\v{S}irca}

\author[JSI]{S.~\v{S}tajner }

\author[USK]{S.~Strauch}

\author[Mainz]{M.~Thiel}

\author[Mainz]{A.~Weber}

\author[TAU]{I.~Yaron\smallskip{}
}

\author{\\\textbf{(A1 Collaboration)}}

\fntext[eq]{These authors contributed equally to this work.}

\cortext[cor2]{Corresponding author}

\address[TAU]{School of Physics and Astronomy, Tel Aviv University, Tel Aviv 69978,
Israel.}

\address[JSI]{Jo\v{z}ef Stefan Institute, 1000 Ljubljana, Slovenia.}

\address[Mainz]{Institut f\"ur Kernphysik, Johannes Gutenberg-Universit\"at, 55099
Mainz, Germany.}

\address[Rutgers]{Rutgers, The State University of New Jersey, Piscataway, NJ 08855,
USA.}

\address[nrc]{Department of Physics, NRCN, P.O. Box 9001, Beer-Sheva 84190, Israel.}

\address[soreq]{Soreq NRC, Yavne 81800, Israel.}

\address[huji]{Racah Institute of Physics, Hebrew University of Jerusalem, Jerusalem
91904, Israel.}

\address[UL]{Faculty of Mathematics and Physics, University of Ljubljana, 1000
Ljubljana, Slovenia.}

\address[USK]{University of South Carolina, Columbia, South Carolina 29208, USA.}
\begin{abstract}
We measured the ratio $P_{x}/P_{z}$ of the transverse to longitudinal
components of polarization transferred from electrons to bound protons
in $^{12}\mathrm{C}$ by the $^{12}\mathrm{C}(\vec{e},e'\vec{p})$
process at the Mainz Microtron (MAMI). We observed consistent deviations
from unity of this ratio normalized to the free-proton ratio, $(P_{x}/P_{z})_{^{12}\mathrm{C}}/(P_{x}/P_{z})_{^{1}\mathrm{H}}$,
for both $s$- and $p$-shell knocked out protons, even though they
are embedded in averaged local densities that differ by about a factor
of two. The dependence of the double ratio on proton virtuality is
similar to the one for knocked out protons from $^{2}\mathrm{H}$
and $^{4}\mathrm{He}$, suggesting a universal behavior. It further
implies no dependence on average local nuclear density.
\end{abstract}

\end{frontmatter}{}

Deviations of quasi-elastic measurements on nuclei from those performed
on protons or from calculations using free-proton form-factors (FFs)
reflect various many-body effects, potentially including medium modifications
of the bound proton structure in the nuclear field~\cite{Dieterich,Strauch}.
The ratio of the transverse ($P_{x}$) to longitudinal ($P_{z}$)
polarization transfer components measured in the elastic double-polarized
process $^{1}\mathrm{H}(\vec{e},e'\vec{p})$ is proportional to the
ratio of the electric to magnetic FFs of the free proton, $R_{^{1}\mathrm{H}}\equiv(P_{x}/P_{z})_{^{1}\mathrm{H}}\propto G_{E}^{p}/G_{M}^{p}$~\cite{Akh74}.
In nuclei, the ratio of the polarization transfer components to a
bound proton, $R_{A}\equiv(P_{x}/P_{z})_{A}$, can be determined from
the analogous quasi-free proton knock-out process $A\left(\vec{e},e'\vec{p}\right)$.
Measurements of $R_{A}$ eliminate many systematic uncertainties and
thus constitute a sensitive and precise tool to study possible deviations
of a bound proton properties from a free one. 

Previous double polarized proton knock-out experiments on light nuclei,
$^{2}\mathrm{H}$ and $^{4}\mathrm{He}$, were found to be in agreement
when compared in terms of the proton virtuality, which is a measure
of the ``off-shellness'' of the bound proton (see Eq.~(\ref{eq:eq_virt})).
The measurements showed no dependence on the average nuclear density
nor on momentum transfer~\cite{deep2012PLB}. For the deuteron, detailed
calculations~\cite{Arenhovel} explained the deviations from the
free proton by final state interactions (FSI). It is thus interesting
to extend the measurements to heavier nuclei were FSI effects are
expected to be different. 

The $^{12}\mathrm{C}$ nucleus is a particularly appealing target
for such studies as one can selectively probe protons from specific
nuclear shells, $s$ and $p$. The average local densities in these
shells differ by about a factor of two, which was predicted to impact
the polarization transfer to $s$- and $p$-shell protons differently~\cite{PhysRevC.87.028202}.
Previous measurements on $s$- and $p$-shell protons in $^{16}\mathrm{O}$
were limited in statistics and the kinematical range covered~\cite{Malov_O16}.

In this paper we report on the measurements of the $P_{x}/P_{z}$
ratio for protons bound in carbon, $^{12}\mathrm{C}\left(\vec{e},e'\vec{p}\right)$,
and present the double ratio $R_{^{12}\mathrm{C}}/R_{^{1}\mathrm{H}}$.
In terms of virtuality, our results exhibit consistency between $s$-
and $p$-shell protons as well as with measurements obtained on other
light nuclei. Thus, they confirm the absence of average nuclear density
dependence even in the heavier nucleus $^{12}\mathrm{C}$.

The experiment was performed at the Mainz Microtron (MAMI) accelerator
using the A1 beam-line and spectrometers~\cite{a1aparatus}. We used
a \SI{600}{MeV} continuous-wave polarized electron beam with a current
of about \SI{10}{\micro A}. The average beam polarization was about
80\%, measured with a M{\o}ller polarimeter and verified by Mott
polarimetry. The uncertainty in the beam polarization was less than
5\%. The beam helicity was flipped at a rate of \SI{1}{Hz}. Two high-resolution,
small solid-angle spectrometers with momentum acceptances of $20-25\,\%$
were used to detect the scattered electrons in coincidence with the
knocked-out protons. The target consisted of three carbon foils of
\SI{0.8}{mm} thickness each, separated by about \SI{1.5}{cm} and
tilted at an angle of \ang{40} with respect to the beam. The usage
of three tilted foils reduced the proton energy loss in the target
and improved the resolution for the reaction-vertex determination.
This reduced the systematic uncertainty in the determined polarization
transfer components at the reaction point. The proton spectrometer
was equipped with a polarimeter placed behind its focal-plane (FPP)
using a \SI{7}{cm} thick carbon analyzer~\cite{a1aparatus,Pospischil:2000pu}.
The spin-dependent scattering of the polarized proton by the carbon
analyzer enables the determination of the proton transverse polarization
components at the focal plane~\cite{Pospischil:2000pu}. The polarization-transfer
components at the reaction point were obtained by correcting the measured
components for the spin precession in the magnetic field of the spectrometer.
Following the convention of~\cite{Strauch}, both $P_{z}$ and $P_{x}$
were determined in the scattering plane, defined by the incident and
scattered electron momenta, where $P_{z}$ is along and $P_{x}$ is
perpendicular to the momentum transfer vector, $\vec{q}$. 

In the analysis, cuts were applied to identify coincident electrons
and protons that originate from the carbon target, and to ensure good
reconstruction of tracks in the spectrometers and the FPP. To remove
Coulomb scattering events by the carbon analyzer, we selected only
events that scattered by more than $\ang{8}$ in the FPP. 
\begin{table}[!t]
\vspace{-0.25cm}
\caption{\label{tab:The-kinematic-settings}The kinematic settings in the experiment.
The angles and momenta represent the central values for the two spectrometers:
$p_{p}$ and $\theta_{p}$ ($p_{e}$ and $\theta_{e}$) are the knocked
out proton (scattered electron) momentum and scattering angles, respectively.}
\medskip{}
\centering{}%
\begin{tabular}{rllcc}
\toprule 
\multicolumn{2}{c}{Kinematic} &  & \multicolumn{2}{c}{Setting}\tabularnewline
\cmidrule{4-5} 
 &  &  & A & B\tabularnewline
\midrule 
$Q^{2}$\negthickspace{}\negthickspace{} & \negthickspace{}\negthickspace{}{[}GeV$^{2}$/c$^{2}${]} &  & 0.40 & 0.18\tabularnewline
$p_{\text{miss}}$\negthickspace{}\negthickspace{} & \negthickspace{}\negthickspace{}{[}MeV/c{]} &  & $-130$ to $100$ & $-250$ to $-100$\tabularnewline
$p_{e}$\negthickspace{}\negthickspace{} & \negthickspace{}\negthickspace{}{[}MeV/c{]} &  & 385 & 368\tabularnewline
$\theta_{e}$\negthickspace{}\negthickspace{} & \negthickspace{}\negthickspace{}{[}deg{]} &  & 82.4 & 52.9\tabularnewline
$p_{p}$\negthickspace{}\negthickspace{} & \negthickspace{}\negthickspace{}{[}MeV/c{]} &  & 668 & 665\tabularnewline
$\theta_{p}$\negthickspace{}\negthickspace{} & \negthickspace{}\negthickspace{}{[}deg{]} &  & $-34.7$ & $-37.8$\tabularnewline
\multicolumn{3}{r}{{\scriptsize{}\# of events after cuts}} & 1.7 M & 1.1 M\tabularnewline
\bottomrule
\end{tabular}
\end{table}

The polarization transfer components $P_{x}$ and $P_{z}$ were first
determined as a function of the proton missing momentum defined as
$\vec{p}_{\text{miss}}=\vec{q}-\vec{p}_{p}$, where $\vec{p}_{p}$
is the outgoing proton momentum. We define the scalar missing momentum,
$p_{\mathrm{miss}}\equiv\pm\left|\vec{p}_{\mathrm{miss}}\right|$,
where the sign is taken to be positive (negative) if the longitudinal
component of $\vec{p}_{\text{miss}}$ is parallel (anti-parallel)
to $\vec{q}$. The measurements were performed in two kinematical
settings that covered two ranges in $p_{\mathrm{miss}}$ and two ranges
in the invariant four-momentum transfer $Q^{2}=\vec{q}^{2}-\omega^{2}$,
where $\omega$ is the energy transfer. Details of the kinematics
are summarized in Table~\ref{tab:The-kinematic-settings}. 
\begin{figure}
\begin{centering}
\includegraphics[width=1\columnwidth]{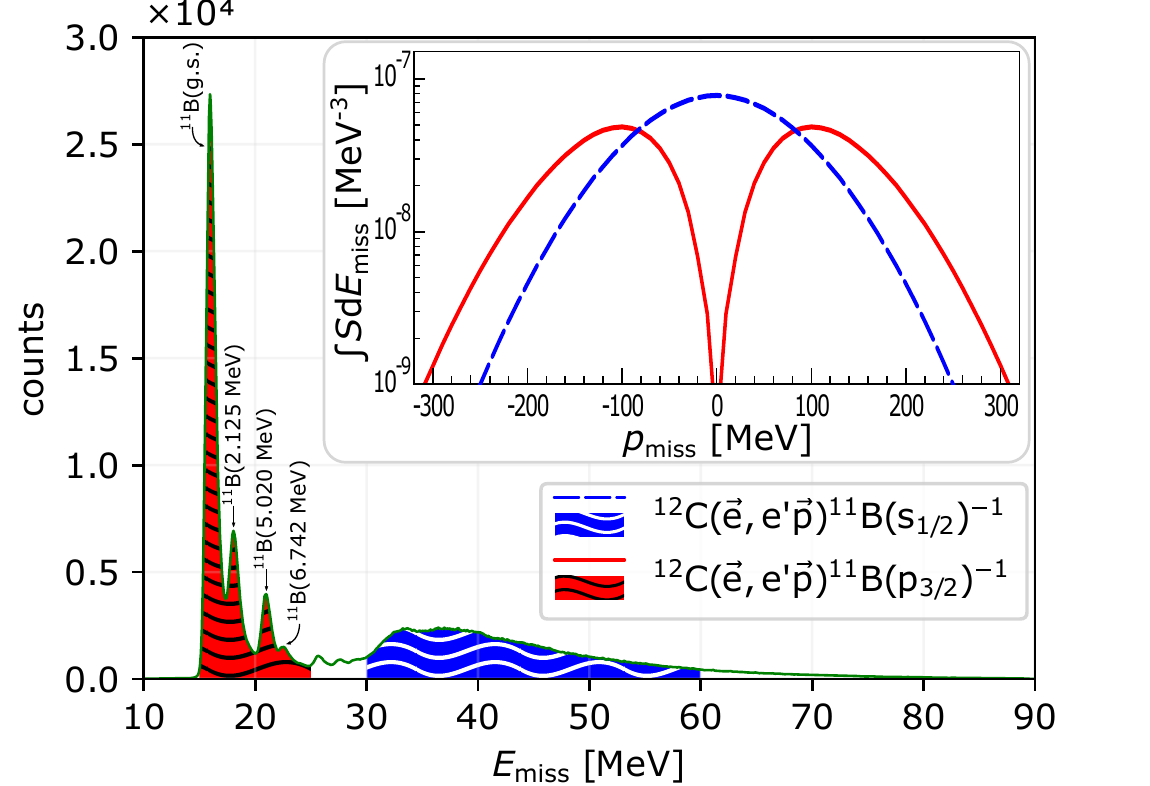}
\par\end{centering}
\caption{\label{fig:fig_emiss}The proton missing energy spectrum for $^{12}$C$\left(e,e'p\right)$
in setup A. The distinct peaks correspond to removal of $p_{3/2}$-shell
protons in $^{12}$C resulting in $^{11}$B ground state and excited
states as noted. The $E_{\text{miss}}$ ranges considered in the analysis
for $p_{3/2}$ and $s_{1/2}$ protons are marked in red and blue,
respectively. The green line denotes the $E_{\text{miss}}$ range
used for the analysis combining all protons. The inset shows the momentum
distribution predictions of the independent particle shell model for
$p_{3/2}$ and $s_{1/2}$ protons in $^{12}$C, obtained from~\cite{Dutta}.}
\end{figure}

The protons knocked out from the $s$ and $p$ shells were identified
by their missing energy. The missing energy is defined as $E_{\text{miss}}\equiv\omega-T_{p}-T_{^{11}\text{B}}$,
where $T_{p}$ is the measured kinetic energy of the outgoing proton,
and $T_{^{11}\text{B}}$ is the calculated kinetic energy of a recoiling
$^{11}$B nucleus (g.s.). The missing-energy spectrum of setting A
is shown in Fig.~\ref{fig:fig_emiss}. The sharp peaks correspond
to the ground state and the lowest excited states of the recoiling
$^{11}$B. Following Dutta \emph{et al}.~\cite{Dutta} we present
the polarization-transfer results for two ranges of $E_{\text{miss}}$
shown in the figure: the first ($15<E_{\text{miss}}<\SI{25}{MeV}$)
corresponds to proton removal primarily from the $^{12}$C $p_{3/2}$
shell; the second ($30<E_{\text{miss}}<\SI{60}{MeV}$) corresponds
predominantly to proton removal from the $s$-shell. The missing energy
cut allows some $s$-shell strength in the $p$-shell region and vice
versa. In addition, we show the combined data from the entire $E_{\text{miss}}$
range ($10<E_{\text{miss}}<\SI{90}{MeV}$) covering proton removal
from both $s$- and $p$-shells. The inset in Fig.~\ref{fig:fig_emiss}
(adapted from~\cite{Dutta}), shows the predicted momentum distributions
of $p$- and $s$-shell protons in $^{12}$C obtained from an independent
particle shell model spectral function (S)~\cite{Dutta}. The difference
between the $s$- and $p$-shell proton momentum distributions around
$p_{\text{miss}}=0$, may impact the polarization transfer in this
region. 

Helicity-independent uncertainties in the measured ratios (acceptance,
detector efficiency, target density, etc.) largely cancel out due
to frequent flips of the beam helicity. The uncertainties in beam
polarization, carbon analyzing power and efficiency are reduced well
below the statistical uncertainty by taking the $P_{x}/P_{z}$ ratio.
The total systematic uncertainty in $R_{^{12}\text{C}}$, dominated
by the vertex position reconstruction in the target, does not exceed
2\% and is about 25\% of the statistical uncertainty. In the following
figures, only the statistical uncertainties are shown.
\begin{figure}[t]
\begin{centering}
\includegraphics[width=0.85\columnwidth]{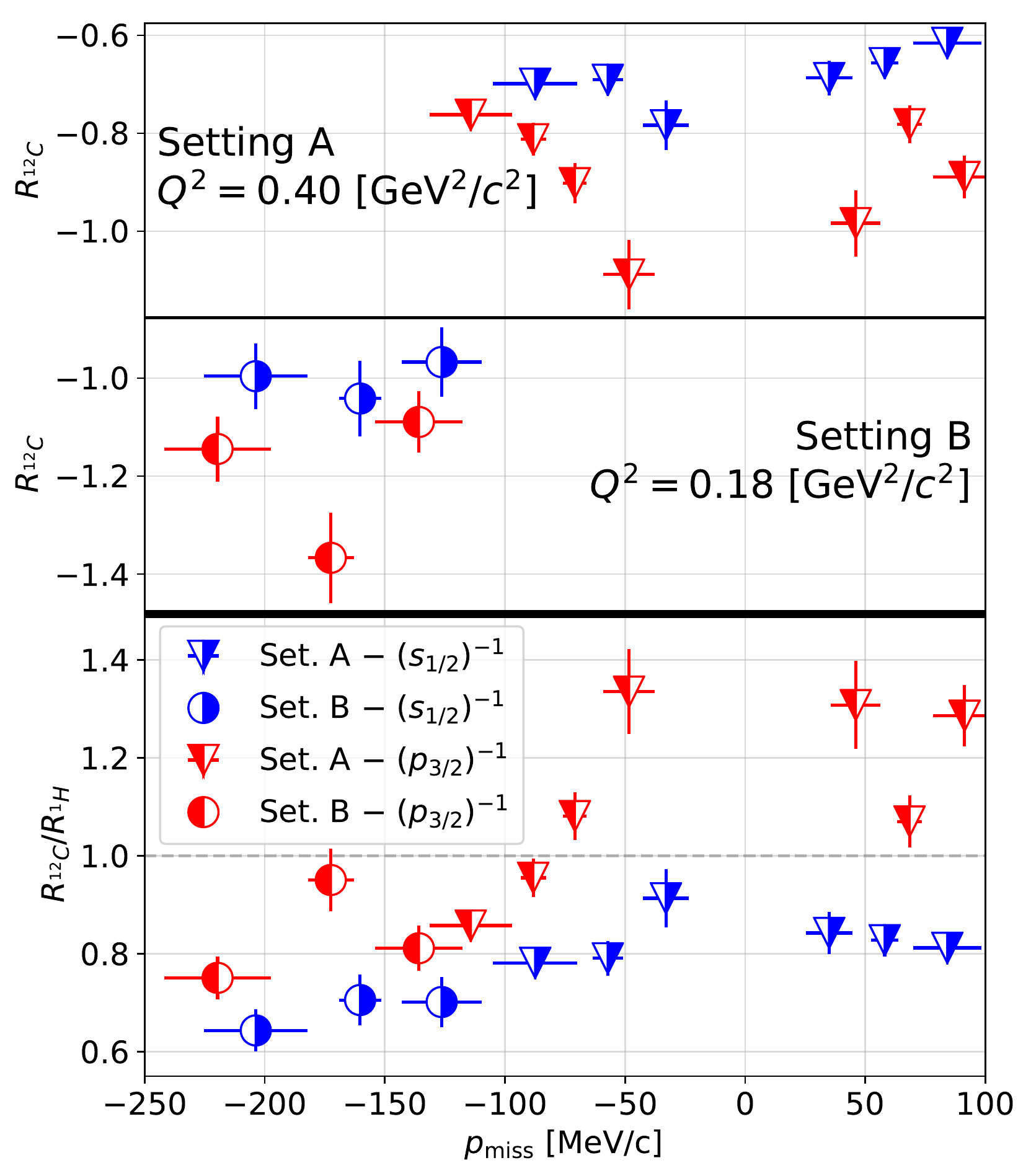}
\par\end{centering}
\caption{\label{fig:cPm}The measured ratio of polarization transfer components,
$R_{^{12}\textrm{C}}$ (top) and the double ratio $R_{^{12}\text{C}}/R_{^{1}\text{H}}$
(bottom), versus the missing-momentum. The measurements for $p_{3/2}$-
and $s_{1/2}$-shell protons are shown separately. The uncertainties
are statistical only, and the horizontal bars indicate the width of
the $p_{\text{miss}}$ distribution in each bin (standard deviation).
The legend is common to all panels in the figure. Note, triangles
(circles) refer to kinematical setting A (B). Symbols open on the
left (right) side refer to $s$- ($p$-) shell removals.}
\end{figure}

The measured helicity-dependent ratios $R_{^{12}\textrm{C}}$ for
both settings are presented in Fig.~\ref{fig:cPm} (top) as a function
of $p_{\mathrm{miss}}$. The difference in $R_{^{12}\textrm{C}}$
between $s$- and $p$-shell proton removal with the same $p_{\mathrm{miss}}$
is clearly visible in the figure. We removed some contributions to
the differences between data at the same $p_{\mathrm{miss}}$, which
are due to the different kinematics (or momentum transfer), by dividing
$R_{^{12}\textrm{C}}$ by the hydrogen ratio \begin{linenomath}
\begin{equation}
R_{^{1}\text{H}}\equiv\left(\frac{P_{x}}{P_{z}}\right)_{^{1}\mathrm{H}}\!\!\!=-\frac{2M_{p}c^{2}}{(E+E^{\prime})\tan\frac{\theta_{e}}{2}}\cdot\frac{G_{E}^{p}(Q^{2})}{G_{M}^{p}(Q^{2})},\label{eq:R_p}
\end{equation}
\end{linenomath} where $E$ is the incident electron energy, and
$M_{p}$ is the proton mass. The scattered electron energy ($E'=E'\left(E,Q^{2}\right)$)
and scattering angle ($\theta_{e}=\theta_{e}\left(E,Q^{2}\right)$)
are calculated assuming elastic electron-proton scattering.

$R_{^{1}\text{H}}$ was calculated on an event by event basis using
the proton FFs parametrized by Bernauer et al.~\cite{Bernauer} and
averaged over the bin. The double-ratio of the $^{12}\text{C}$ data
to $^{1}\mathrm{H}$, $R_{^{12}\text{C}}/R_{^{1}\text{H}}$, is shown
in the bottom panel of Fig.~\ref{fig:cPm}. However, even after division
by $R_{^{1}\text{H}}$, the differences between $s$- and $p$-shell
results at the same missing momenta are still significant.

\begin{figure}
\begin{centering}
\vspace{-1pt}
\includegraphics[width=1\columnwidth]{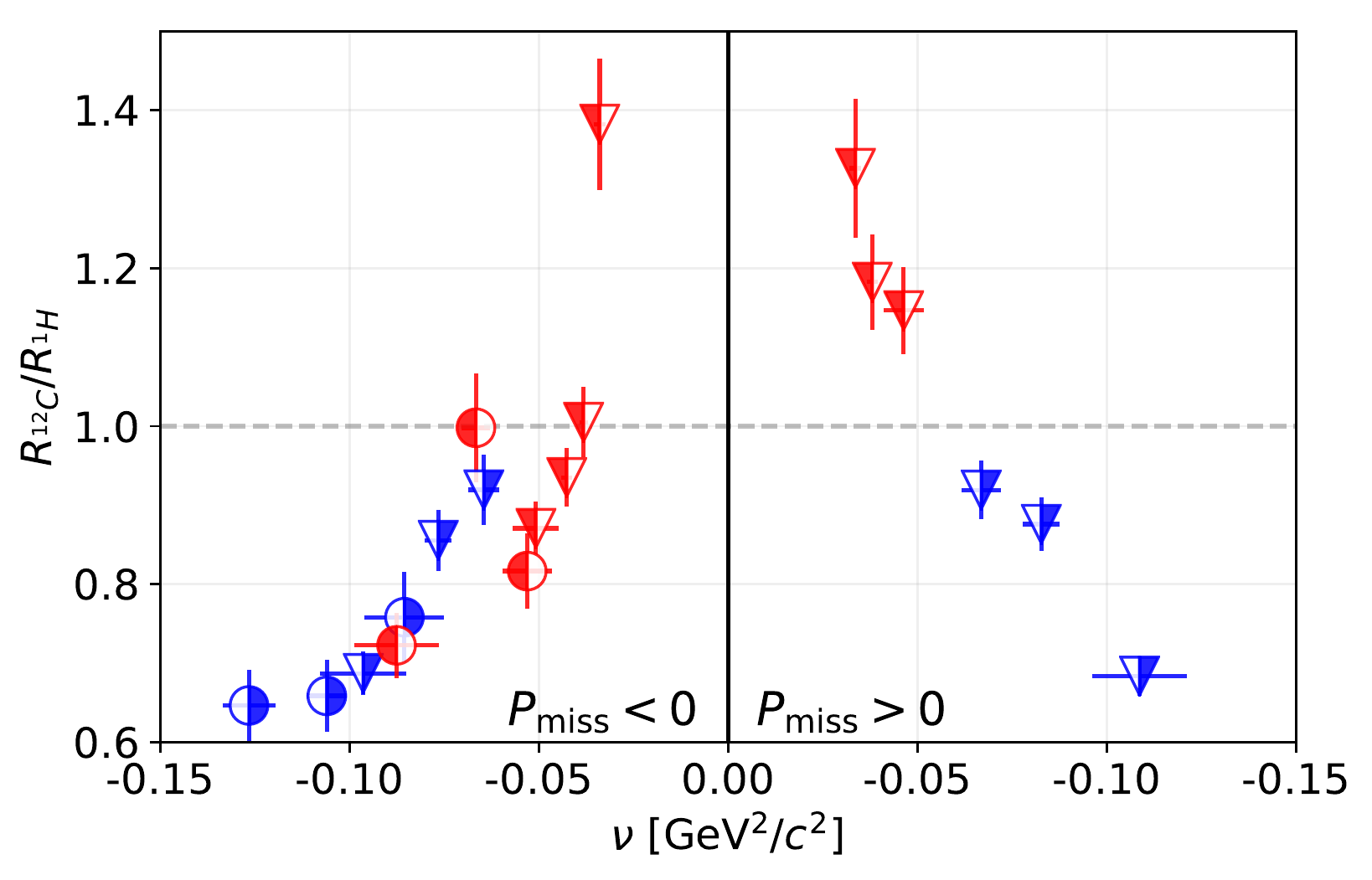}
\par\end{centering}
\caption{\label{fig:fig_cvirt}The measured double-ratio, $R_{^{12}\text{C}}/R_{^{1}\text{H}}$,
for $p$- and $s$-shell protons, as a function of the proton virtuality.
The virtuality dependence is shown separately for positive and negative
missing momenta. We used the same symbols as in Fig.~\ref{fig:cPm}.}
\end{figure}

The bound nucleon can be characterized also by its virtuality, i.e.
its ``off-shellness''. There is no unique way to define virtuality.
Following~\cite{deep2012PLB} we define the virtuality, $\nu$, of
a bound proton as \begin{linenomath}
\begin{equation}
\nu\!\equiv\!\Big(M_{A}c\!-\!\sqrt{M_{A-1}^{2}c^{2}\!+\!p_{\text{miss}}^{2}}\Big)^{2}\!-\!p_{\text{miss}}^{2}\!-\!M_{p}^{2}c^{2},\label{eq:eq_virt}
\end{equation}
\end{linenomath}where $M_{A}$ is the mass of the target nucleus,
$M_{A-1}\equiv\sqrt{\left(\omega-E_{p}+M_{A}\right)^{2}-p_{\mathrm{miss}}^{2}}$
(determined event by event), and $E_{p}$ is the total energy of the
outgoing proton. We note that the virtuality (Eq.~(\ref{eq:eq_virt}))
is not a unique function of $p_{\mathrm{miss}}$. This is demonstrated
in the two dimensional event distribution of $\nu$ vs $p_{\mathrm{miss}}$
shown in the supplementary material~\cite{supplemental}. Equation~(\ref{eq:eq_virt})
implies that the struck proton is off-shell ($\ne M_{p}^{2}c^{2}$)
and the recoil system is on-shell. The virtuality dependence of $R_{^{12}\text{C}}/R_{^{1}\text{H}}$
is shown in Fig.~\ref{fig:fig_cvirt}. The double ratios are shown
separately for positive and negative missing momenta due to possible
differences, as observed in $^{4}\mathrm{He}$~\cite{Strauch} and
calculated for $^{2}\mathrm{H}$ due to FSI~\cite{deep2012PLB}. 

The $s$- and $p$-shell protons have different wave functions as
reflected also in their missing-momentum distributions. These differences,
such as the behavior at $p_{\text{miss}}=0$ (see~\cite{Dutta} and
Fig.~\ref{fig:fig_emiss}) and possibly the total angular momentum,
may affect the polarization transfer, as predicted by calculations~\cite{Ryckebusch99,GiustiCarlotta}.
Nevertheless, the corresponding double-ratios have the same smooth
behavior, and show the same virtuality dependence, as is clearly shown
in Fig.~\ref{fig:fig_cvirt}. Motivated by the observed good agreement
between the $s$- and $p$-shell protons as a function of virtuality,
we combined the data and obtained $R_{^{12}\text{C}}/R_{^{1}\text{H}}$
for the entire missing energy region, $10<E_{\text{miss}}<\SI{90}{MeV}$.
\begin{figure*}[!t]
\begin{centering}
\includegraphics[width=0.8\textwidth]{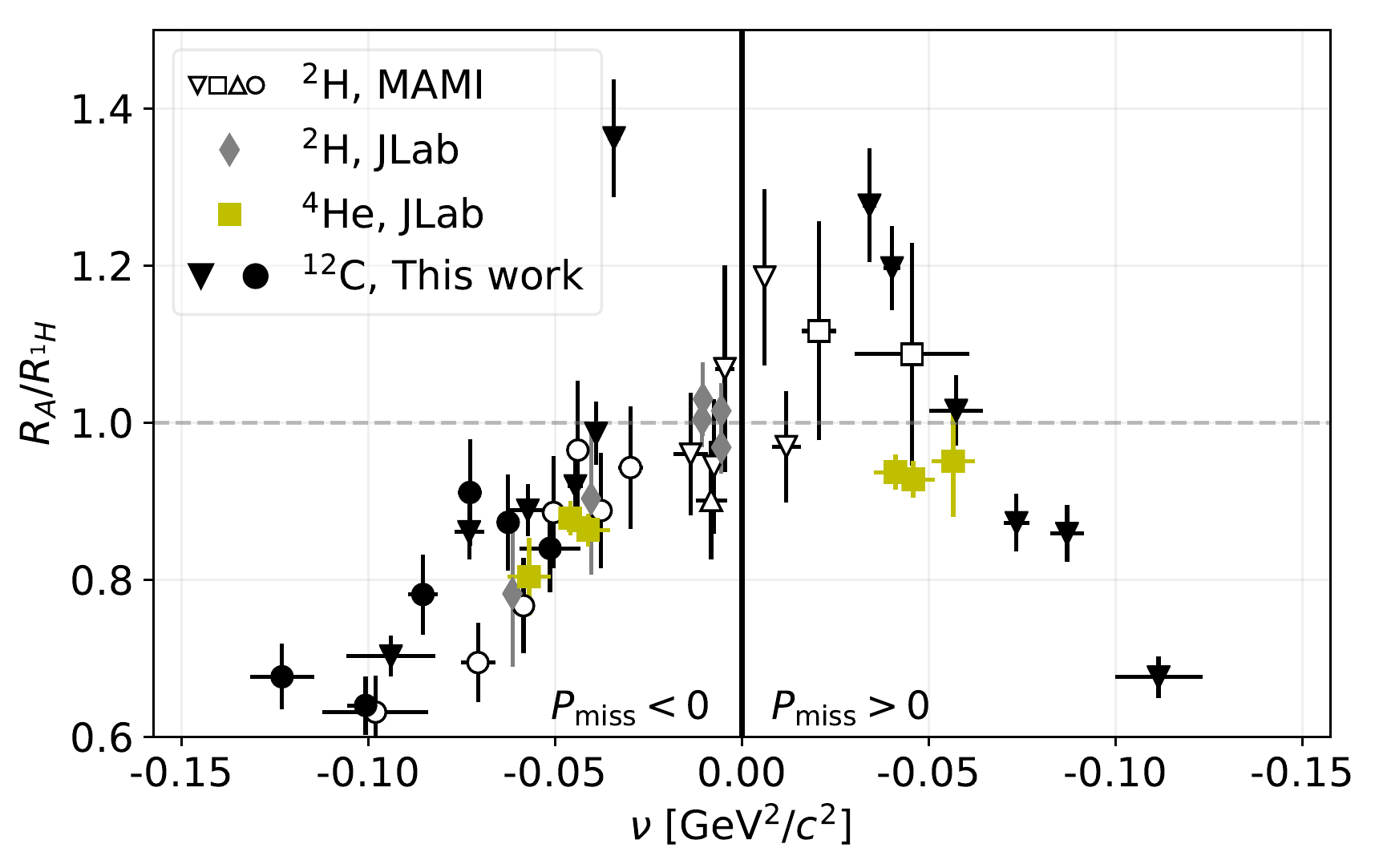}
\par\end{centering}
\caption{\label{fig:fig_univ}The measured double-ratio of protons from $^{12}\text{C}$
compared with those obtained for $^{2}$H and $^{4}$He ($R_{A}/R_{^{1}\text{H}}$,
$A=$ $^{2}$H, $^{4}$He, $^{12}$C) as a function of the proton
virtuality. The $^{12}\text{C}$ data (black full symbols) are the
combined $s$- and $p$-shell removal of this work. The open symbols
are $^{2}$H data measured at Mainz~\cite{deep2012PLB} with essentially
the same kinematics as the present work ~\cite{deep2012PLB}. The
light symbols are $^{2}$H and $^{4}$He data measured at JLab at
$Q^{2}=\text{\SIlist[list-units=single]{1;0.8}{GeV\squared}}/c^{2}$,
respectively~\cite{jlabDeep,Strauch}. The triangles (circles) refer
to setting A (B) as in Fig.~\ref{fig:cPm}.}
\end{figure*}

In Fig.~\ref{fig:fig_univ}, the $^{12}\text{C}$ double ratios,
combined for $s$- and $p$-shell proton removals, are compared with
those of $^{2}$H obtained at MAMI for the same kinematics~\cite{deep2012PLB},
as well as to $^{2}$H and $^{4}$He data measured at JLab at $Q^{2}=\text{\SIlist[list-units=single]{1;0.8}{GeV\squared}}/c^{2}$,
respectively~\cite{jlabDeep,Strauch}. Note that the data shown in
Fig.~\ref{fig:fig_univ} are not identical to those in Fig.~\ref{fig:fig_cvirt}
due to the different $E_{\text{miss}}$ range and bins. The new $^{12}$C
data almost double the virtuality range covered by the data from light
nuclei. The higher values of $R_{^{12}\text{C}}/R_{^{1}\text{H}}$
at $\left|\nu\right|<\SI{0.04}{GeV\squared}/c^{2}$ are due to $p_{3/2}$
protons whose behavior is attributed to the $p$ wave function properties
at small $\left|p_{\mathrm{miss}}\right|$~\cite{Ryckebusch99,GiustiCarlotta},
unlike $s$-shell protons in the other nuclei.

The data suggest that the double-ratio is characterized well by the
virtuality of the struck proton. Virtuality seems to be a better parameter
than $p_{\mathrm{miss}}$ to describe polarization transfer to a struck
proton in different nuclei.

To test this hypothesis, we compared the $^{12}$C data to the MAMI
$^{2}$H data. To enable an event-by-event comparison, we adjusted
the theoretical calculations~\cite{Arenhovel} for $^{2}\mathrm{H}$
to reproduce the measured polarization transfer components of $^{2}$H.
The ratio of the $^{12}$C data to the adjusted model is $1.07\pm0.03$.
In this comparison we excluded the afore mentioned $\left|\nu\right|<\SI{0.04}{GeV\squared}/c^{2}$
data, due to its special behavior. 

The agreement between the data from the different nuclei suggests
that the observed deviations from the free proton ratio have a common
origin. We note that Eq.~(\ref{eq:R_p}) is valid only for a free
proton. Thus the double ratio in nuclei does not remove FSI effects.
When the deuteron data~\cite{deep2012PLB} were compared to the calculations
of~\cite{Arenhovel}, the theory was in good agreement with the data.
This implies that in the deuteron most of the deviation is due to
FSI. One may speculate that the same effects dominate in heavier nuclei
as well, although this has to be confirmed by detailed calculations.
These should take into account the variation of the kinematics over
the experimental phase space. Such a task is rather involved and may
depend on various parameters. The new $^{12}$C data may provide important
information for determination of the mechanisms at work and the validity
of using free proton FFs in such calculations. 

The data confirm that the virtuality of the struck proton is the preferred
parameter for comparing the deviations from a free proton. This implies
that further measurements to look for local nuclear density effects
should compare polarization transfer to protons at different local
densities, like different shells, but with the same virtuality as
can be deduced from Fig.~\ref{fig:fig_cvirt}. This comparison requires
high statistics in order to confirm or reject in medium modification
of the proton FFs, which are estimated at a few percent~\cite{PhysRevC.87.028202}.

To summarize, our data of the polarization-transfer ratios for $^{12}$C
extend the previous nuclear measurements on $^{2}$H and almost double
the virtuality range. The new double ratios $R_{^{12}\mathrm{C}}/R_{^{1}\mathrm{H}}$
agree well with those previously measured on $^{2}$H and $^{4}$He,
including those obtained in different kinematics. The double-ratios
exhibit a similar shape for nuclei with very different average local
density. The new results suggest also that measurements of both $^{2}$H
and $^{4}$He over an extended virtuality range are needed. Indeed,
such measurements were proposed~\cite{jlabfutureprop} and approved
at JLab.

We would like to thank the Mainz Microtron operators and technical
crew for the excellent operation of the accelerator. This work is
supported by the Israel Science Foundation (Grant 390/15) of the Israel
Academy of Arts and Sciences, by the Israel Ministry of Science, Technology
and Space, by the Deutsche Forschungsgemeinschaft with the Collaborative
Research Center 1044, by the U.S. National Science Foundation (PHY-1205782),
and by the Croatian Science Foundation Project No.~1680. We acknowledge
the financial support from the Slovenian Research Agency (research
core funding No.~P1\textendash 0102).\newpage{}

\bibliographystyle{elsarticle-num}
\addcontentsline{toc}{section}{\refname}\bibliography{ceep048}
\clearpage{}

\newpage{}\newpage{}\newpage{}

\includepdf[pages=-,width=0.9\paperwidth,height=1\paperheight,keepaspectratio]{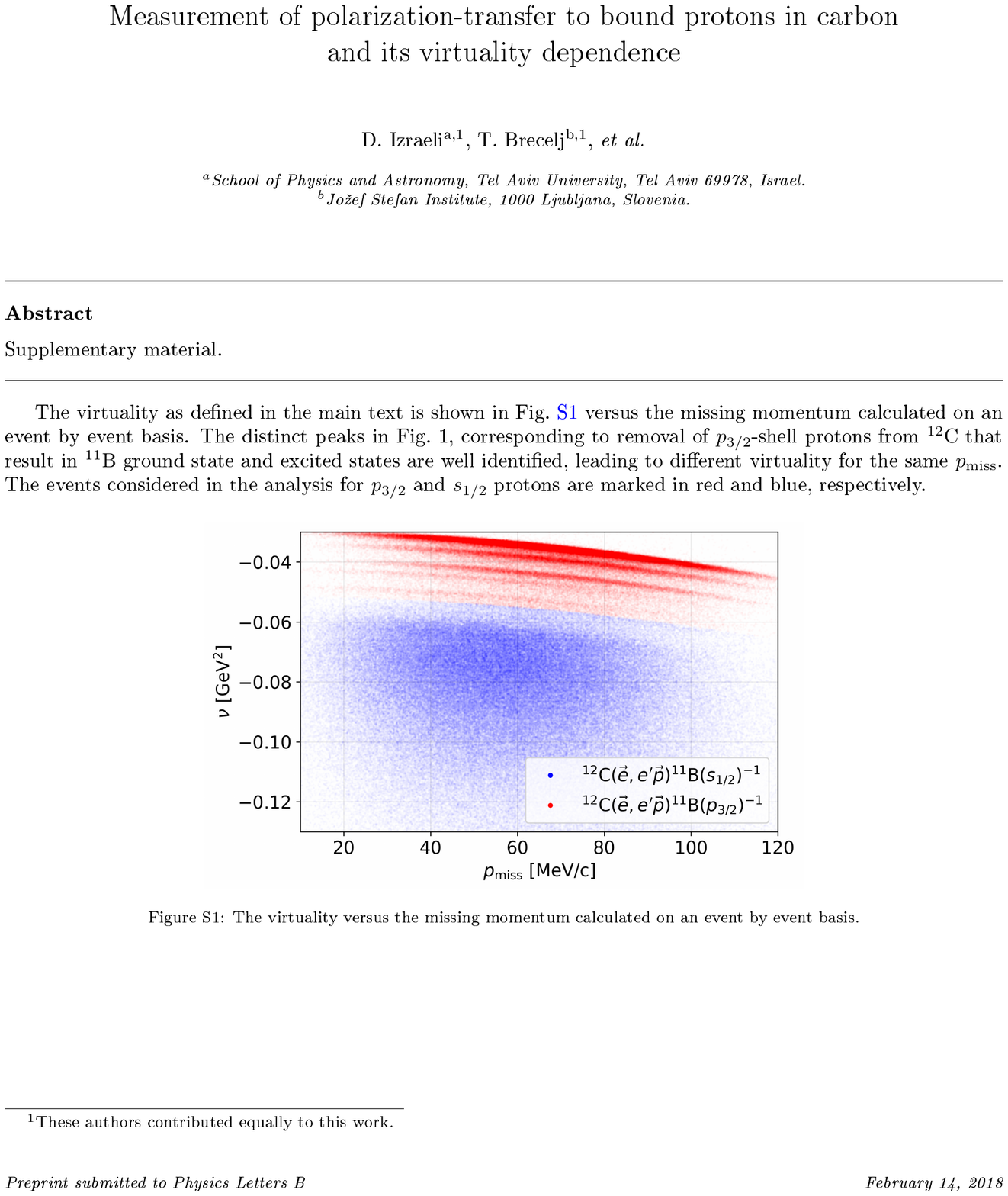}
\end{document}